\documentclass[aps,twocolumn,superscriptaddress,prd]{revtex4-1}

\usepackage{bm,graphicx,textcomp,amssymb,amsmath,dcolumn,color}

\usepackage{slashed}

\usepackage{multirow}
\usepackage{tabularx}

\usepackage[utf8]{inputenc}
\usepackage[T1]{fontenc}

\newcommand{\tr}{\operatorname{tr}}

\newcommand{\nn}{\nonumber\\}

\newcommand{\f}[1]{\mbox{\boldmath$#1$}}

\newcommand{\na}{\mbox{\boldmath$\nabla$}}
\newcommand{\bea}{\begin{eqnarray}}
\newcommand{\ea}{\end{eqnarray}}
\newcommand{\eea}{\end{eqnarray}}
\newcommand{\ord}{\,{\cal O}}

\usepackage{ulem}
\usepackage{comment}
\usepackage{bbm}
\usepackage{bbold}   
\usepackage{mathtools}

\usepackage{tikz-feynman}
\tikzfeynmanset{compat=1.1.0} 
\usepackage{tikz} 
\usepackage{ctable}

\begin{document}

\title{Field-assisted birefringent Compton scattering}

\author{N.~Ahmadiniaz} 
\affiliation{Helmholtz-Zentrum Dresden-Rossendorf, 
Bautzner Landstra{\ss}e 400, 01328 Dresden, Germany,} 
\author{T.~E.~Cowan}
\affiliation{Helmholtz-Zentrum Dresden-Rossendorf, 
Bautzner Landstra{\ss}e 400, 01328 Dresden, Germany,} 
\affiliation{Institut f\"ur Kern- und Teilchenphysik, Technische Universit\"at Dresden, 01062 Dresden, Germany,}

\author{M.~Ding} 
\affiliation{Helmholtz-Zentrum Dresden-Rossendorf, 
Bautzner Landstra{\ss}e 400, 01328 Dresden, Germany,} 

\author{M.~A.~Lopez Lopez} 
\affiliation{Helmholtz-Zentrum Dresden-Rossendorf, 
Bautzner Landstra{\ss}e 400, 01328 Dresden, Germany,} 

\author{R.~Sauerbrey} 
\affiliation{Helmholtz-Zentrum Dresden-Rossendorf, 
Bautzner Landstra{\ss}e 400, 01328 Dresden, Germany,} 

\author{R.~Shaisultanov}
\affiliation{Helmholtz-Zentrum Dresden-Rossendorf, 
Bautzner Landstra{\ss}e 400, 01328 Dresden, Germany,}
\affiliation{ELI Beamlines Centre, Institute of Physics, Czech Academy of Sciences, Za Radnic\`{i} 835, 25241 Doln\`{i} B\v{r}e\v{z}any, Czech Republic,}

\author{R.~Sch\"utzhold}
\affiliation{Helmholtz-Zentrum Dresden-Rossendorf, 
Bautzner Landstra{\ss}e 400, 01328 Dresden, Germany,}
\affiliation{Institut f\"ur Theoretische Physik, 
Technische Universit\"at Dresden, 01062 Dresden, Germany,}

\begin{abstract}
Motivated by experimental initiatives such as the Helmholtz International 
Beamline for Extreme Fields (HIBEF), we study Compton scattering of x-rays 
at electrons in a strong external field (e.g., a strong optical laser) with 
special emphasis on the polarization-changing (i.e., birefringent)  
contribution on the amplitude level. 
Apart from being a potential background process for the planned vacuum 
birefringence experiments, this effect could be used for diagnostic purposes. 
Since the birefringent signal from free electrons 
(i.e., without the external field) vanishes in forward direction, the ratio 
of the birefringent and the normal (polarization conserving) contribution
yields information about the field strength at the interaction point. 
\end{abstract}

\date{\today} 

\maketitle

\section{Introduction} 

Besides the color and intensity of light, its polarization can also be used to 
obtain information about the viewed object. 
As a well known example, by placing a transparent object such as a plastic 
ruler between two polarization filters, one may visualize the lines of 
tension in the material.
Modern technology allows us to transfer this concept to different regimes, 
for example the investigation of plasmas or even the vacuum with polarized 
x-rays on ultra-short time and length scales. 

In the following, we investigate one of the most important elementary 
processes in this respect -- the scattering of x-ray photons at electrons, 
usually referred to as Compton or Thomson scattering. 
Of course these processes have already been studied in 
numerous works theoretically and experimentally, see, e.g., \cite{thomson06,vach62,brown64,sarach70,esar93,hart98,johnson12,chen98,
compton23-1,compton23-2,jaun23,compton24,kron26,mandl52,bloch34,eisen70,
rib75,ehlo78,akh84,ehlo87,ehlo89,berg93,
cooper85,alt04,eisaman11,krafft10,pratt10,grie12,kraj12,seipt14,bula96,
bamber99,cole18,liu14,
silva20,venk20,dipiazza18,ilderton19,mack16,mack19,acosta20,heinzl20,seipt20,
dipiazza21,king21,ahre17,black18,shai19,ilderton20,kamp21,gelfer22,ahmadi20}.
Here, as motivated above, we place special emphasis on the polarization 
changing (i.e., birefringent) signal in or close to forward direction.  
Furthermore, in order to determine whether the scattering contributions from 
many electrons add up coherently or incoherently, we consider the scattering
amplitudes (instead of probabilities or cross sections) and we do not average 
over the electron spins. 

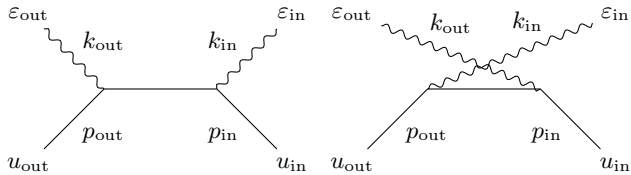
\begin{figure}[ht]
\centering
\begin{minipage}[ht]{0.2\textwidth}
	\begin{tikzpicture}[scale=1.0] 
	\begin{feynman}
	\vertex (a) at (0,0);
	\vertex (b) at (1.5,0);
	\vertex (i1) at (-1,-1) {\(u_{\text{out}}\)};
	\vertex (f1) at (-1,1) {\(\varepsilon_{\text{out}}\)};
	\vertex (f2) at (2.5,-1) {\(u_{\text{in}}\)};
	\vertex (i2) at (2.5,1) {\(\varepsilon_{\text{in}}\)};
	\diagram*{  (i1)  -- [edge label'=\(p_{\text{out}}\)] (a) -- [photon,edge label'=\(k_{\text{out}}\)] (f1) };
	\diagram*{	(a) -- [plain] (b)};
	\diagram*{	(i2)  -- [photon,edge label'=\(k_{\text{in}}\)] (b) --  [edge label'=\(p_{\text{in}}\)] (f2) };
	\end{feynman}
	\end{tikzpicture}	
\end{minipage}
\hspace{0.5cm}
\begin{minipage}[ht]{0.2\textwidth}
	\centering
	\begin{tikzpicture}[scale=1.0]
	\begin{feynman}
	\vertex (a) at (0,0);
	\vertex (b) at (1.5,0);
	\vertex (i1) at (-1,-1) {\(u_{\text{out}}\)} ;
	\vertex (f1) at (-1,1) {\(\varepsilon_{\text{out}}\)};
	\vertex (i2) at (2.5,-1) {\(u_{\text{in}}\)};
	\vertex (f2) at (2.5,1)  {\(\varepsilon_{\text{in}}\)};
	\diagram* {(i1) -- [edge label'=\(p_{\text{out}}\)] (a) -- [draw=none] (f1) };	
	\diagram* {	(a) -- [ edge label'=\( \)] (b)};
	\diagram* {	(f2) --  [draw=none] (b)	-- [edge label'=\(p_{\text{in}}\)] (i2) };
	\diagram* {
		(a) -- [photon, edge label=\(k_{\text{in}}\),near end] (f2),
		(f1) -- [photon, edge label=\(k_{\text{out}}\),near start] (b),
	};
	\end{feynman}
	\end{tikzpicture}	
\end{minipage}
	
\caption{Lowest-order Feynman diagrams for Thomson or Compton scattering 
of free electrons.}
\label{fig1} 
\end{figure}

For free electrons as in Fig.~\ref{fig1}, 
the birefringent signal vanishes in forward direction. 
Thus, we consider electrons under the influence of a strong external field 
such as an optical laser, where the birefringent signal in forward direction 
(in comparison to the normal, i.e., polarization conserving, contribution)
allows us to infer the strength of the external field.
Apart from this diagnostic purpose, the birefringent scattering of x-rays at 
electrons represents an important background process in planned 
experiments aiming at the detection of quantum vacuum birefringence,
especially for the Coulomb-assisted case proposed in \cite{ahmadi21}. 
Thus, investigating this process is also important for estimating the 
background in these experiments. 

\section{Free Electrons}\label{Free Electrons}

Let us first revisit the well-known case of free electrons, where the two 
lowest-order Feynman diagrams are depicted in Fig.~\ref{fig1}. 
The associated amplitude reads ($\hbar=c=1$)
\bea
\label{amplitude-free}
{\mathfrak A}
=
\bar u_{\rm out} 
{\mathcal M}
u_{\rm in} 
\,,
\ea
where $u_{\rm in}$ and $u_{\rm out}$ are the initial and final electron 
spinors while the matrix ${\mathcal M}$ contains the Dirac structure 
\bea
\label{matrix-free}
{\cal M}
=
\slashed{\varepsilon}_{\rm in}
\frac{q^2}{\slashed{p}_{\rm in}-\slashed{k}_{\rm out}-m}
\slashed{\varepsilon}_{\rm out}
+
\slashed{\varepsilon}_{\rm out}
\frac{q^2}{\slashed{p}_{\rm in}+\slashed{k}_{\rm in}-m}
\slashed{\varepsilon}_{\rm in}
\quad
\ea
in terms of the initial and final (linear) photon polarization vectors 
$\varepsilon_{\rm in} $ and $\varepsilon_{\rm out}$,
photon momenta $k_{\rm in} $ and $k_{\rm out}$, as well as  
electron momenta $p_{\rm in} $ and $p_{\rm out}$.

Using transversality 
$k_{\rm in}\cdot\varepsilon_{\rm in}=k_{\rm out}\cdot\varepsilon_{\rm out}=0$
and the temporal gauge 
$p_{\rm in}\cdot\varepsilon_{\rm in}=p_{\rm in}\cdot\varepsilon_{\rm out}=0$ 
with respect to the initial electron rest frame as well as 
$(\slashed{p}_{\rm in}-m)u_{\rm in}=0$, this matrix ${\mathcal M}$ 
can be simplified to 
\bea
\label{matrix-free-simplify}
{\cal M}
=
-\frac{q^2}{2}
\left(
\frac{\slashed{\varepsilon}_{\rm in}\slashed{\varepsilon}_{\rm out}
\slashed{k}_{\rm out}}{p_{\rm in}\cdot k_{\rm out}}
+
\frac{\slashed{\varepsilon}_{\rm out}\slashed{\varepsilon}_{\rm in}
\slashed{k}_{\rm in}}{p_{\rm in}\cdot k_{\rm in}}
\right)
\,.
\ea
Very often the next step would be to consider the absolute value squared 
$|{\mathfrak A}|^2$ and to average over the initial and final electron 
spins (e.g., via the Casimir trick) and possibly even over the photon  
polarizations. 
However, here we are interested in the amplitudes and their dependence 
on electron spin and photon polarization. 

Quite generally, an elegant way to proceed in this case 
is to split the matrix $\mathcal M$
into its scalar, pseudo-scalar, vector, axial-vector, and tensor part via 
\bea
\label{split}
{\mathcal M}=
{\mathbb S}
+{\mathbb P}\gamma^5
+{\mathbb V}_\mu\gamma^\mu
+{\mathbb A}_\mu\gamma^\mu\gamma^5
+{\mathbb T}_{\mu\nu}\Sigma^{\mu\nu}
\,,
\ea
where $\gamma^5=i\gamma^0\gamma^1\gamma^2\gamma^3$ and 
$\Sigma^{\mu\nu}=i[\gamma^\mu,\gamma^\nu]/4$, see \cite{kirvo94}.
As is well known, the 16 matrices $\bf{1}$, $\gamma^5$, $\gamma^\mu$,
$\gamma^\mu\gamma^5$, and $\Sigma^{\mu\nu}$ form a complete basis set 
which is orthogonal with respect to the trace of their products. 
Thus, the coefficients can be obtained by multiplying the matrix $\mathcal M$
with the corresponding basis element and taking the trace 
${\mathbb S}=\tr\{{\mathcal M}\}/4$ and  
${\mathbb P}=\tr\{{\mathcal M}\gamma^5\}/4$ etc. 

For the matrix~\eqref{matrix-free-simplify}, we find 
${\mathbb S}={\mathbb P}={\mathbb T}_{\mu\nu}=0$ and 
\bea
\label{free-v}
{\mathbb V}^\mu
&=&
\frac{q^2}{2}
\left[
\frac{k_{\rm out}\cdot\varepsilon_{\rm in} }{p_{\rm in}\cdot k_{\rm out}  }
\,\varepsilon_{\rm out}^\mu 
+
\frac{k_{\rm in}\cdot\varepsilon_{\rm out} }{p_{\rm in}\cdot k_{\rm in}  }
\,\varepsilon_{\rm in}^\mu 
\right.
\nn
&&
\left. 
- 
\varepsilon_{\rm in}\cdot\varepsilon_{\rm out}
\left(
\frac{k_{\rm out}^\mu }{p_{\rm in}\cdot k_{\rm out}  }
+
\frac{k_{\rm in}^\mu }{p_{\rm in}\cdot k_{\rm in}  }
\right) 
\right] 
\ea
for the vector part, while the axial-vector term reads 
\bea
\label{free-a}
{\mathbb A}_\mu
=i\epsilon_{\mu\nu\rho\sigma} 
\,\frac{q^2}{2}
\left(
\frac{k_{\rm out}^\nu \varepsilon_{\rm in}^\rho\varepsilon_{\rm out}^\sigma}
{p_{\rm in}\cdot k_{\rm out}  }
-
\frac{k_{\rm in}^\nu \varepsilon_{\rm in}^\rho\varepsilon_{\rm out}^\sigma }
{p_{\rm in}\cdot k_{\rm in}  }
\right) 
\,.
\ea
At this point, we may already draw several important conclusions. 
Due to the anti-symmetry of the Levi-Civita symbol 
$\epsilon_{\mu\nu\rho\sigma}$, the axial-vector term ${\mathbb A}_\mu$
describes birefringent scattering with 
$\varepsilon_{\rm in}\neq\varepsilon_{\rm out}$. 
Furthermore, this contribution vanishes in forward direction 
$k_{\rm in}=k_{\rm out}$. 
More precisely, in view of momentum conservation $\Delta\f{p}=-\Delta\f{k}$, 
this birefringent matrix element scales with $(\Delta\f{k})^2$ such that the 
probability or differential cross section behaves as $(\Delta\f{k})^4$. 
As another point, the matrix elements 
$\bar u_{\rm out}\gamma^\mu\gamma^5u_{\rm in}$
strongly depend on the electron spins. 
More precisely, after averaging over the electron spins, 
the mean values of the amplitudes vanish. 
%
%
Thus, for unpolarized electrons, their scattering amplitudes 
would typically not add up coherently.

In contrast, the vector term ${\mathbb V}^\mu$ does contain 
polarization conserving scattering with 
$\varepsilon_{\rm in}=\varepsilon_{\rm out}$. 
Actually, in forward direction $k_{\rm in}=k_{\rm out}$, 
only polarization conserving scattering is possible due to 
${\mathbb V}^\mu\propto\varepsilon_{\rm in}\cdot\varepsilon_{\rm out}$. 
The vector term ${\mathbb V}^\mu$ does also contain the classical limit: 
For $\mu=0$, the matrix element $\bar u_{\rm out}\gamma^\mu u_{\rm in}$ 
simplifies to $u_{\rm out}^\dagger u_{\rm in}$ and is thus independent of 
the electron spins.
As a result, the scattering amplitudes of many electrons would add up 
coherently, even if they are unpolarized. 

This fact should not be too surprising as this term ${\mathbb V}^0$ 
reproduces the classical limit of Thomson scattering.
After transforming to the initial rest frame of the electrons 
$p_{\rm in}=(m,0,0,0)$, we find 
\bea
\label{Thomson}
{\mathbb V}^0=-q^2\frac{\varepsilon_{\rm in}\cdot\varepsilon_{\rm out}}{m} 
\,,
\ea
which is just the amplitude for Thomson scattering.

\section{Field-assisted scattering} 

After these preliminaries, let us study electrons under the influence of 
an external laser field.
Assuming that the field is not too strong (see below), we consider the 
six lowest-order Feynman diagrams depicted in Fig.~\ref{fig2}
where $\varepsilon_{\rm L}$ and $k_{\rm L}$ denote polarization and 
momentum of the laser photons. 
Using the same decomposition as in Eq.~\eqref{split}, we again find 
${\mathbb S}={\mathbb P}={\mathbb T}_{\mu\nu}=0$. 
%
%
In principle, the remaining terms ${\mathbb V}^\mu$ and ${\mathbb A}^\mu$
can again be obtained via multiplying $\mathcal M$ with $\gamma^\mu$ and 
$\gamma^\mu\gamma^5$, respectively, and taking the trace. 
However, the resulting expressions are too long for a discussion of the 
general case. 

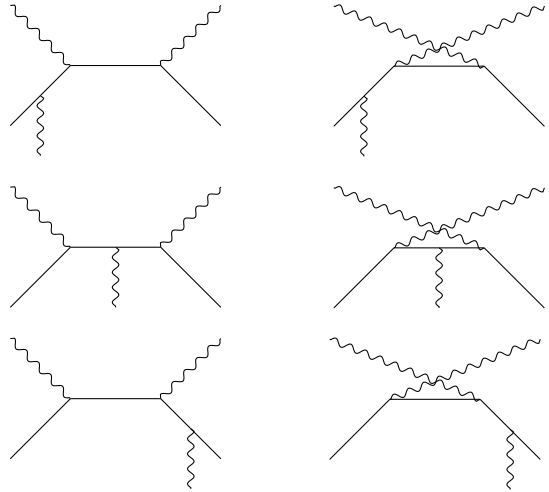
\begin{figure}[ht]

\begin{center}
	\begin{minipage}{0.2\textwidth}		
		\begin{tikzpicture}[scale=0.8] 
		\begin{feynman}		
		\vertex (a) at (0,0);
		\vertex (c) at (-0.5,-0.5);
		\vertex (b) at (1.5,0);
		\vertex (i1) at (-1,-1);
		\vertex (f1) at (-1,1);
		\vertex (i2) at (2.5,1);
		\vertex (f2) at (2.5,-1) ;
		
		\diagram* {(a) --  (b), (i1)  -- (c)-- (a) -- [photon] (f1), (i2) -- [photon] (b) --  (f2) };
		\vertex (d) at (-0.5,-1.5) ;
		\diagram*{(d) -- [photon] (c)};
		\end{feynman}
		\end{tikzpicture}
	\end{minipage}
\hspace{0.5cm}
	\begin{minipage}{0.2\textwidth}
		\begin{tikzpicture}[scale=0.8] 
		\begin{feynman}
		\vertex (a) at (0,0);
		\vertex (c) at (-0.5,-0.5);
		\vertex (b) at (1.5,0);
		\vertex (i1) at (-1,-1) ;
		\vertex (f1) at (-1,1);
		\vertex (i2) at (2.5,-1);
		\vertex (f2) at (2.5,1) ;
		\diagram* {(i1) 	--   (c)-- (a) -- [draw=none] (f1) ,	(a) --  (b),(f2) 	--  [draw=none] (b)	--  (i2) };
		\vertex (d) at (-0.5,-1.5) ;
		\diagram* {(d) -- [photon] (c) ,(a) -- [photon ] (f2),(f1) -- [photon ] (b)};
		\end{feynman}
		\end{tikzpicture}
	\end{minipage}	
\end{center}	
\begin{center}
	\begin{minipage}{0.2\textwidth}
		\begin{tikzpicture}[scale=0.8] 
		\begin{feynman}
		\vertex (a) at (0,0);
		\vertex (c) at (0.75,0);
		\vertex (b) at (1.5,0);
		\vertex (i1) at (-1,-1);
		\vertex (f1) at (-1,1) ;
		\vertex (i2) at (2.5,1) ;
		\vertex (f2) at (2.5,-1) ;	
		\diagram*{
			(a) -- (c) --  (b),
			(i1)  --  (a) -- [photon] (f1) ,
			(i2)  -- [photon] (b) --  (f2),		
		};
		
		\vertex (d) at (0.75,-1) ;
		\diagram*{
			(d) -- [photon] (c) ,
		};
		\end{feynman}
		\end{tikzpicture}
	\end{minipage}
\hspace{0.5cm}
	\begin{minipage}{0.2\textwidth}
		
		\begin{tikzpicture}[scale=0.8] 
		\begin{feynman}
		\vertex (a) at (0,0);
		\vertex (c) at (0.75,0);
		\vertex (b) at (1.5,0);
		\vertex (i1) at (-1,-1) ;
		\vertex (f1) at (-1,1) ;
		\vertex (i2) at (2.5,-1);
		\vertex (f2) at (2.5,1) ;
		
		\diagram*{
			(a) --  (c) -- (b),
			(i1) 	--  (a) -- [draw=none] (f1) ,	
			(f2) 	--  [draw=none] (b)	--  (i2) ,
			
		};
		
		\vertex (d) at (0.75,-1) ;
		\diagram* {
			(d) -- [photon] (c) ,
			(a) -- [photon] (f2),
			(f1) -- [photon] (b),
		};
		\end{feynman}
		\end{tikzpicture}
		
	\end{minipage}
	
\end{center}
\begin{center}
	\begin{minipage}{0.2\textwidth}
		
		\begin{tikzpicture}[scale=0.8] 
		\begin{feynman}
		
		\vertex (a) at (0,0);
		\vertex (c) at (2.0,-0.5);
		\vertex (b) at (1.5,0);
		\vertex (i1) at (-1,-1) ;
		\vertex (f1) at (-1,1) ;
		\vertex (i2) at (2.5,1) ;
		\vertex (f2) at (2.5,-1) ;
		
		\diagram* {
			(a) --  (b),
			(i1)  --   (a) -- [photon] (f1) ,
			(i2)  -- [photon] (b)-- (c) --  (f2) ,
		};
		
		\vertex (d) at (2.0,-1.5);
		\diagram*{
			(d) -- [photon] (c),
		};
		
		\end{feynman}
		\end{tikzpicture}
	\end{minipage}
\hspace{0.5cm}
	\begin{minipage}{0.2\textwidth}

		\begin{tikzpicture}[scale=0.8] 
		\begin{feynman}
		\vertex (a) at (0,0);
		\vertex (c) at (2.0,-0.5);
		\vertex (b) at (1.5,0);
		\vertex (i1) at (-1,-1) ;
		\vertex (f1) at (-1,1) ;
		\vertex (i2) at (2.5,-1) ;
		\vertex (f2) at (2.5,1);
		\diagram* {
			(i1) 	--   (a) -- [draw=none] (f1) ,	
			(a) --  (b),
			(f2) 	--  [draw=none] (b)	-- (c)--  (i2) ,
		};
		\vertex (d) at (2.0,-1.5) ;
		\diagram* {
			(d) -- [photon] (c) ,
			(a) -- [photon ] (f2),
			(f1) -- [photon ] (b),
		};
		\end{feynman}
		\end{tikzpicture}
		\label{fig2}
	\end{minipage}
	
\end{center}

\caption{Lowest-order Feynman diagrams for Thomson or Compton scattering 
of electrons under the influence of an external laser field which is 
represented by the insertion of an additional photon line.}
\label{fig2} 
\end{figure}

Thus, let us simplify the expressions by specifying them to relevant 
example cases. 
Without loss of generality (w.l.o.g.) we may transform to the initial 
rest frame of the electron $p_{\rm in}=(m,0,0,0)$.
Then, as motivated above, we focus on the birefringent signal in forward 
direction, i.e., we may set (again w.l.o.g.) 
\bea
\label{wlog}
k_{\rm in} &=& \omega_{\rm in}(1,1,0,0)
\nn
k_{\rm out} &=& \omega_{\rm out}(1,1,0,0)
\nn
\varepsilon_{\rm in} &=& (0,0,1,0)
\nn
\varepsilon_{\rm out} &=& (0,0,0,1)
\,.
\ea
Note that, in contrast to the case of free electrons, forward scattering 
does not imply $k_{\rm in}=k_{\rm out}$ due to the energy $\omega_{\rm L}$
of the additional laser photon. 
In view of the hierarchy $m\gg\omega_{\rm in}\gg\omega_{\rm L}$, we find 
$\omega_{\rm out}\approx\omega_{\rm in}\pm\omega_{\rm L}$, depending on 
whether the laser photon represents an incoming or an outgoing line. 
Here, we focus on the case 
$\omega_{\rm out}\approx\omega_{\rm in}+\omega_{\rm L}$,
but the other one 
$\omega_{\rm out}\approx\omega_{\rm in}-\omega_{\rm L}$
can be treated in complete analogy. 
The remaining momentum transfer $\Delta\f{p}=\ord(\omega_{\rm L})$
is taken up by the electron with small recoil energy 
$\ord(\omega_{\rm L}^2/m)$. 

\subsection{Faraday scenario}\label{Faraday scenario} 

After fixing the coordinate system according to Eq.~\eqref{wlog}, 
we have to specify the vectors $\varepsilon_{\rm L}$ and $k_{\rm L}$
associated to the optical laser. 
For optical materials, a well-known effect is Faraday rotation 
describing the change of light polarization due to a magnetic field 
oriented along the propagation axis. 
To study the analogue of this effect in our scenario, let us align 
the magnetic field of the optical laser with the x-ray propagation 
direction, i.e., the $x$ axis. 
Then, the optical laser must propagate in an orthogonal direction, 
so let us choose the $y$ axis such that 
\bea
\label{Faraday}
k_{\rm L} &=& \omega_{\rm L}(1,0,1,0)
\nn
\varepsilon_{\rm L} &=& (0,0,0,1)
\,.
\ea
In this scenario, we find ${\mathbb A}^\mu=0$ and 
\bea
\label{Faraday-v}
{\mathbb V}^\mu
=
\frac{q^3B_{\rm L}}{2\omega_{\rm in}(m+\omega_{\rm L})m}\,
(2,-1,-1,0)
\,,
\ea
where $B_{\rm L}$ denotes the magnetic field strength of the optical laser. 
As in the previous Section~\ref{Free Electrons}, the term ${\mathbb V}^0$
yields the dominant contribution and reproduces the classical limit 
(discussed in the next Section).

Comparing Eqs.~\eqref{Thomson} and \eqref{Faraday-v}, we find that 
the leading-order ratio between the birefringent ${\mathfrak A}_\perp$
and the normal (polarization conserving) amplitude ${\mathfrak A}_\|$ 
scales with the inverse of the combined Keldysh parameter 
\bea
\label{Keldysh}
\frac{{\mathfrak A}_\perp}{{\mathfrak A}_\|}
\approx 
\frac{qB_{\rm L}}{m\omega_{\rm in}}
\ea
which compares the strength $B_{\rm L}$ of the optical laser 
with the x-ray frequency $\omega_{\rm in}$.
Since this ratio~\eqref{Keldysh} is typically a small number, 
birefringent scattering is suppressed. 

\subsection{Cotton-Mouton scenario}\label{Cotton-Mouton scenario}

In order to suppress birefringent scattering even further, let us consider 
another scenario where the x-ray propagates perpendicularly to the magnetic 
field of the optical laser, for example parallel to its electric field 
\bea
\label{Cotton-Mouton}
k_{\rm L} &=& \omega_{\rm L}(1,0,1,0)
\nn
\varepsilon_{\rm L} &=& (0,1,0,0)
\,.
\ea
In this case, there is no ${\mathbb V}^0$ term 
\bea
\label{Cotton-Mouton-v}
{\mathbb V}^\mu
=
\frac{q^3B_{\rm L}}{2\omega_{\rm in}(m+\omega_{\rm L})m}\,
(0,0,0,1)
\,,
\ea
but we also get an axial-vector contribution 
\bea
\label{Cotton-Mouton-a}
{\mathbb A}^\mu
=
\frac{q^3B_{\rm L}}{2\omega_{\rm in}(m+\omega_{\rm L})m}\,
(-i,0,0,0)
\,. 
\ea
Even though the pre-factor is the same as in Eq.~\eqref{Faraday-v}, 
the matrix structure is very different.
The matrix $\mathcal M$ determining the amplitude 
${\mathfrak A}=\bar u_{\rm out}{\mathcal M}u_{\rm in}$ 
only contains off-diagonal blocks and thus only products 
between the large and the small components of the bi-spinors 
$\bar u_{\rm out}$ and $u_{\rm in}$ contribute, leading to 
an additional suppression by $\omega_{\rm L}/m$ in view of 
the momentum transfer $\Delta\f{p}=\ord(\omega_{\rm L})$.
Thus, the birefringent amplitude is much stronger suppressed 
in this case 
\bea
\label{Cotton-Mouton-suppressed}
\frac{{\mathfrak A}_\perp}{{\mathfrak A}_\|}
=
\ord\left(  
\frac{qB_{\rm L}}{m\omega_{\rm in}}\,
\frac{\omega_{\rm L}}{m}
\right) 
\,.
\ea
In addition, the matrix elements determining the amplitude 
${\mathfrak A}=\bar u_{\rm out}{\mathcal M}u_{\rm in}$ 
depend on the electron spins. 
Thus, for unpolarized electrons, their amplitudes would not add 
up coherently -- in contrast to the ${\mathbb V}^0$ contribution 
in the Faraday scenario -- which implies further suppression in 
the case of many electrons. 

However, one should remember that this estimate~\eqref{Cotton-Mouton-suppressed}
is based on the six lowest-order Feynman diagrams in Fig. \ref{fig2}, 
which include one vertex with the external field. 
Since $\omega_{\rm L}/m$ is such a small number, it is possible that 
higher-order Feynman diagrams including two or more vertices with the 
external laser field, which would scale with $\ord(q^4B^2_{\rm L})$ or higher, 
yield birefringent amplitudes ${\mathfrak A}_\perp$ larger than the one above 
-- provided that the laser field $B_{\rm L}$ is strong enough, i.e., 
$qB_{\rm L}>\omega_{\rm in}\omega_{\rm L}$. 

\section{Classical Scattering} 

Let us compare the results of the full quantum calculation above 
(though only to lowest order) with the corresponding classical picture. 
First, we start with the non-relativistic equation of motion for the 
electron 
\bea
\label{eom-classical}
m\ddot{\f{r}}=q\left(\f{E}+\dot{\f{r}}\times\f{B}\right) 
\,.
\ea
As usual the scattering approach, we split the full electromagnetic field 
$\f{E}$ and $\f{B}$ into the strong external background field 
$\f{E}_0$ and $\f{B}_0$ plus the weaker contributions 
$\f{E}_1$ and $\f{B}_1$ from the x-ray 
\bea
\label{split-classical}
\f{E}&=&\f{E}_0+\f{E}_1
\,,
\nn
\f{B}&=&\f{B}_0+\f{B}_1
\,,
\ea
and analogously for the electron trajectory 
\bea
\label{trajectory-classical}
\f{r}&=&\f{r}_0+\f{r}_1
\,,
\ea
where $\f{r}_0$ is supposed to be a solution to the equation of 
motion~\eqref{eom-classical} in the background field 
$\f{E}_0$ and $\f{B}_0$.
Then we may linearize the full equation of motion~\eqref{eom-classical}
in the first-order quantities $\f{E}_1$, $\f{B}_1$, and $\f{r}_1$
\bea
\label{first-order-classical}
m\ddot{\f{r}}_1
&=&
q\left(\f{E}_1+\dot{\f{r}}_1\times\f{B}_0+
\dot{\f{r}}_0\times\f{B}_1\right) 
\nn
&&
+q\f{r}_1\cdot\na\left(\f{E}_0+\dot{\f{r}}_0\times\f{B}_0\right) 
\,.
\ea
The solution of this linear equation for $\f{r}_1(t)$ then allows us to 
infer the emitted (i.e., scattered) radiation. 

\subsection{Magnetic Field} 

As motivated by the Faraday scenario in Sec.~\ref{Faraday scenario}, 
let us first consider the case of a pure magnetic field, i.e., $\f{E}_0=0$, 
which we take to be constant for simplicity $\f{B}_0=\rm const$. 
This case admits the simple zeroth-order solution $\f{r}_0=0$ and thus we 
find 
\bea
\label{magnetic-classical}
m\ddot{\f{r}}_1
=
q\left(\f{E}_1+\dot{\f{r}}_1\times\f{B}_0\right) 
\,.
\ea
In principle, for a driving term $q\f{E}_1$ which is harmonically 
oscillating with the x-ray frequency $\omega_{\rm in}$, 
this set of linear equations can be solved exactly. 
However, in order to compare the classical approach to the results 
of the previous Section, we use a perturbative solution strategy. 

Assuming that $qB_0/(m\omega_{\rm in})$ is small, cf.~Eq.~\eqref{Keldysh}, 
we may expand the solution $\f{r}_1$ of Eq.~\eqref{magnetic-classical}  
in terms of this small parameter. 
The zeroth-order solution 
\bea
\label{zeroth-order-classical}
{\f{r}}_1=-\frac{q\f{E}_1}{m\omega^2_{\rm in}} 
\left[
1+\ord\left(\frac{qB_0}{m\omega_{\rm in}}\right) 
\right]
\ea
just oscillates in the same direction as the incoming x-ray field $\f{E}_1$
and thus emits radiation with the same polarization. 
As a result, it corresponds to the normal (polarization conserving) 
amplitude ${\mathfrak A}_\|$ from Eq.~\eqref{Thomson}, i.e., 
Thomson scattering.   

The next order, on the other hand,  
\bea
\label{first-order-classical-magnetic}
m\delta\ddot{\f{r}}_1
=
q\dot{\f{r}}_1\times\f{B}_0
\;\leadsto\; 
\delta{\f{r}}_1\propto 
\frac{q^2\f{E}_1\times\f{B}_0}{m^2\omega^3_{\rm in}} 
\ea
describes oscillations orthogonal to the incoming x-ray field $\f{E}_1$
which lead to the emission of radiation with the other polarization. 
Ergo, it yields the birefringent amplitude ${\mathfrak A}_\perp$
to lowest order and thus the ratio of the birefringent and normal signals 
reads (to lowest order)
\bea
\label{Faraday-classical}
\frac{{\mathfrak A}_\perp}{{\mathfrak A}_\|}=
\frac{q\f{e}_{\rm in}\times\f{B}_0}{m\omega_{\rm in}}\cdot\f{e}_{\rm out}
\,,
\ea
with the initial and final x-ray polarizations 
$\f{e}_{\rm in}$ and $\f{e}_{\rm out}$. 

With this simple classical picture, we can already explain the leading-order 
result~\eqref{Keldysh} of the Faraday scenario in Sec.~\ref{Faraday scenario}.
For the Cotton-Mouton scenario in Sec.~\ref{Cotton-Mouton scenario}, 
the above expression~\eqref{Faraday-classical} vanishes -- which means 
that we have to include higher orders in $qB_0/(m\omega_{\rm in})$.
One way to obtain them could be to iterate the above procedure via 
\bea
\label{iterate-classical}
\f{r}_1
&=&
-\frac{q\f{E}_1}{m\omega_{\rm in}^2}
+\frac{q^2\dot{\f{E}}_1\times\f{B}_0}{m^2\omega_{\rm in}^4}
\nn
&&
+\frac{q^3([\f{E}_1+\dot{\f{r}}_1\times\f{B}_0]\times\f{B}_0)\times\f{B}_0}
{m^3\omega_{\rm in}^4}
\,.
\ea
Even if the magnetic field is orthogonal to the x-ray propagation direction 
such that $(\f{e}_{\rm in}\times\f{B}_0)\cdot\f{e}_{\rm out}=0$, 
the second-order term $(\f{E}_1\times\f{B}_0)\times\f{B}_0$ from the 
second line of Eq.~\eqref{iterate-classical} could still yield a non-vanishing 
contribution. 
In this case, the birefringent amplitude scales as 
\bea
\label{Cotton-Mouton-classical}
\frac{{\mathfrak A}_\perp}{{\mathfrak A}_\|}=
\frac{q^2([\f{e}_{\rm in}\times\f{B}_0]\times\f{B}_0)\cdot\f{e}_{\rm out}}
{m^2\omega_{\rm in}^2}
\,.
\ea
In the quantum calculation, this term would correspond to Feynman diagrams 
including two vertices with the external field, which is the reason why 
we do not see this contribution in Eqs.~\eqref{Cotton-Mouton-v} and 
\eqref{Cotton-Mouton-a}.
Still, as already mentioned in Sec.~\ref{Cotton-Mouton scenario}, this 
second-order contribution~\eqref{Cotton-Mouton-classical} could exceed 
to first-order term~\eqref{Cotton-Mouton-suppressed}.
In this case, if the goal is to suppress birefringent scattering from 
electrons as much as possible, one could eliminate this 
contribution~\eqref{Cotton-Mouton-classical} by aligning the 
magnetic field $\f{B}_0$ with the polarization direction $\f{e}_{\rm in}$
of the incoming x-rays. 

\subsection{General Case} 

To conclude this Section, let us briefly discuss the other terms in 
Eq.~\eqref{first-order-classical}.
In principle, the terms including $\dot{\f{r}}_0$ can be eliminated by 
transforming to the (instantaneous) rest frame of the electrons. 
However, this is only a valid approximation if the change of the velocity 
$\dot{\f{r}}_0$ due to the acceleration by the external field 
$\f{E}_0$ and $\f{B}_0$ can be neglected during the scattering process. 
Note that the relevant figure of merit is again the combined Keldysh 
parameter $qE_0/(m\omega_{\rm in})$ as in Eq.~\eqref{Keldysh}, which 
measures how much the electron is accelerated during one x-ray period. 

Assuming that this parameter is small enough 
and transforming to the electron rest frame, the gradient term 
$q(\f{r}_1\cdot\na)\f{E}_0$ remains. 
For an optical laser, this gradient is very small $\ord(\omega_{\rm L})$ 
but for the Coulomb fields of nuclei, for example, the gradients can be 
much larger if the distances between the electrons and the nuclei are small 
enough. 
Since the polarization of the emitted radiation stemming from this term 
is set by the direction of the Coulomb field $\f{E}_0$ at the electron 
position instead of the incoming x-ray field $\f{E}_0$, 
this contribution can also lead to birefringent scattering. 
However, if the electrons are randomly distributed around the nuclei,
their scattering amplitudes would not add up coherently. 

%
%
%

\section{Conclusions} 

We studied the Compton scattering of x-rays at electrons with special emphasis
on the birefringent signal in forward direction. 
Furthermore, in order to determine whether the scattering amplitudes from many 
(typically unpolarized) electrons would add up coherently or not, we consider 
amplitudes instead of probabilities (or cross sections) and their dependence
on the electrons spins. 
An elegant way to do this is to split the Dirac matrix into its 
scalar ${\mathbb S}$, 
pseudo-scalar ${\mathbb P}$,
vector ${\mathbb V}_\mu$, 
axial-vector ${\mathbb A}_\mu$, and 
tensor ${\mathbb T}_{\mu\nu}$
contributions. 
For the cases considered here only the vector ${\mathbb V}_\mu$ and 
axial-vector ${\mathbb A}_\mu$ parts are non-zero. 

For free electrons, there is no birefringent scattering in forward 
direction where only the vector term ${\mathbb V}_\mu$ survives. 
The axial-vector part ${\mathbb A}_\mu$ would describe birefringent 
scattering, but it vanishes in forward direction.
This absence of birefringence in forward direction can be illuminated 
via the following intuitive picture.
If birefringent scattering was present, it would also be possible to 
transform an incoming left-circularly polarized photon into an outgoing 
right-circularly polarized one. 
This would correspond to a change in angular momentum of $2\hbar$ 
(oriented along the propagation direction). 
Since a flip of the internal electron spin $s=\pm\hbar/2$ can only account 
for half of that value, the rest should be compensated by orbital angular
momentum.
However, in forward direction, there is no momentum transfer to the (free) 
electron and thus this is not possible. 

For electrons under the influence of an external field, on the other hand, 
this picture changes as the scattering process can now involve at least one 
additional laser photon, which carries extra angular momentum and leads to 
a non-vanishing momentum transfer (electron recoil), even in forward direction. 
Consistent with this picture, we find birefringent scattering in forward 
direction for electrons within an external field, i.e., an optical laser.  

For the Faraday scenario where the external magnetic field is aligned with 
the x-ray propagation direction, the axial-vector ${\mathbb A}_\mu$ part 
vanishes.  
The vector contribution ${\mathbb V}_\mu$ remains non-zero and does describe 
birefringent scattering. 
As a result, comparing normal (polarization-conserving) and birefringent 
scattering yields information about the external field strength -- 
which could be exploited for diagnostic purposes. 
Note that the amount of birefringent scattering is not governed by the 
Keldysh or laser parameter $qE_{\rm L}/(m\omega_{\rm L})$ of the optical 
laser itself but rather by the combined Keldysh or laser parameter
$qE_{\rm L}/(m\omega_{\rm in})$ which compares the strength of the 
optical laser with the x-ray frequency.  

For the Cotton-Mouton scenario where the x-ray propagates orthogonal to 
the external magnetic field, the first-order birefringent scattering 
amplitude is further suppressed by an additional factor of $\omega_{\rm L}/m$. 
Furthermore, there is another important difference to the Faraday case:
As for free electrons, the ${\mathbb V}_0$ term corresponds to the 
classical limit in the Faraday scenario.
Hence the leading-order amplitude does not depend on the electron 
spins and thus the amplitudes of many electrons can add up coherently.
In contrast, the leading-order amplitude in the Cotton-Mouton scenario
has vanishing ${\mathbb V}_0$ and does depend on the electron spins -- 
resulting in an incoherent superposition (for unpolarized electrons). 

\acknowledgments 

We would like to thank C. Schubert, C. Kohlf\"urst, D. Seipt and  U. Hernandez Acosta for valuable discussions. 
R. S. acknowledges support by the Deutsche Forschungsgemeinschaft 
(DFG, German Research Foundation) -- Project-ID 278162697-- SFB 1242. M. D. is grateful for support by Helmholtz-Zentrum Dresden-Rossendorf High Potential Program. 


\end{document}